\documentclass[aps,prl,preprint,nopacs,superscriptaddress]{revtex4}

\usepackage{graphicx}
\usepackage{verbatim}
\usepackage{mathrsfs}
\usepackage{amsmath,amsfonts,amssymb}
\usepackage{graphicx}

\def\3{2.8in}    
\def\2{2.5in}
\def\4{3.0in}

\def \beq {\begin{equation}}
\def \eeq {\end{equation}}
\begin{document}

\title{Spin polarization and texture of the Fermi arcs in the Weyl Fermion semimetal TaAs}

\author{Su-Yang Xu$^*$}\affiliation {Laboratory for Topological Quantum Matter and Spectroscopy (B7), Department of Physics, Princeton University, Princeton, New Jersey 08544, USA}
\author{Ilya Belopolski$^*$}\affiliation {Laboratory for Topological Quantum Matter and Spectroscopy (B7), Department of Physics, Princeton University, Princeton, New Jersey 08544, USA}
\author{Daniel S. Sanchez$^*$}\affiliation {Laboratory for Topological Quantum Matter and Spectroscopy (B7), Department of Physics, Princeton University, Princeton, New Jersey 08544, USA}

\author{Madhab Neupane\footnote{These authors contributed equally to this work.}}\affiliation {Laboratory for Topological Quantum Matter and Spectroscopy (B7), Department of Physics, Princeton University, Princeton, New Jersey 08544, USA}\affiliation {Condensed Matter and Magnet Science Group, Los Alamos National Laboratory, Los Alamos, NM 87545, USA}
\author{Guoqing Chang}\affiliation{Centre for Advanced 2D Materials and Graphene Research Centre National University of Singapore, 6 Science Drive 2, Singapore 117546}\affiliation{Department of Physics, National University of Singapore, 2 Science Drive 3, Singapore 117542}

\author{Koichiro Yaji}\affiliation{The Institute for Solid State Physics (ISSP), University of Tokyo, Kashiwa, Chiba 277-8581, Japan}

\author{Zhujun Yuan}\affiliation{International Center for Quantum Materials, School of Physics, Peking University, China}
\author{Chenglong Zhang}\affiliation{International Center for Quantum Materials, School of Physics, Peking University, China}
\author{Kenta Kuroda}\affiliation{The Institute for Solid State Physics (ISSP), University of Tokyo, Kashiwa, Chiba 277-8581, Japan}

\author{Guang Bian}\affiliation{Laboratory for Topological Quantum Matter and Spectroscopy (B7), Department of Physics, Princeton University, Princeton, New Jersey 08544, USA}
\author{Cheng Guo}\affiliation{International Center for Quantum Materials, School of Physics, Peking University, China}

\author{Hong Lu}\affiliation{International Center for Quantum Materials, School of Physics, Peking University, China}

\author{Tay-Rong Chang}\affiliation{Department of Physics, National Tsing Hua University, Hsinchu 30013, Taiwan}

\author{Nasser Alidoust}\affiliation {Laboratory for Topological Quantum Matter and Spectroscopy (B7), Department of Physics, Princeton University, Princeton, New Jersey 08544, USA}
\author{Hao Zheng}\affiliation {Laboratory for Topological Quantum Matter and Spectroscopy (B7), Department of Physics, Princeton University, Princeton, New Jersey 08544, USA}

\author{Chi-Cheng Lee}
\affiliation{Centre for Advanced 2D Materials and Graphene Research Centre National University of Singapore, 6 Science Drive 2, Singapore 117546}
\affiliation{Department of Physics, National University of Singapore, 2 Science Drive 3, Singapore 117542}
\author{Shin-Ming Huang}
\affiliation{Centre for Advanced 2D Materials and Graphene Research Centre National University of Singapore, 6 Science Drive 2, Singapore 117546}
\affiliation{Department of Physics, National University of Singapore, 2 Science Drive 3, Singapore 117542}

\author{Chuang-Han Hsu}\affiliation{Centre for Advanced 2D Materials and Graphene Research Centre National University of Singapore, 6 Science Drive 2, Singapore 117546}
\affiliation{Department of Physics, National University of Singapore, 2 Science Drive 3, Singapore 117542}

\author{Horng-Tay Jeng}
\affiliation{Department of Physics, National Tsing Hua University, Hsinchu 30013, Taiwan}
\affiliation{Institute of Physics, Academia Sinica, Taipei 11529, Taiwan}

\author{Arun Bansil}\affiliation{Department of Physics, Northeastern University, Boston, Massachusetts 02115, USA}
\author{Aris Alexandradinata}\affiliation{Department of Physics, Yale University, New Haven, Connecticut 06520, USA}
\author{Titus Neupert}\affiliation {Joseph Henry Laboratory, Department of Physics, Princeton University, Princeton, New Jersey 08544, USA}

\author{Fumio Komori}\affiliation{The Institute for Solid State Physics (ISSP), University of Tokyo, Kashiwa, Chiba 277-8581, Japan}

\author{Takeshi Kondo}\affiliation{The Institute for Solid State Physics (ISSP), University of Tokyo, Kashiwa, Chiba 277-8581, Japan}
\author{Shik Shin}\affiliation{The Institute for Solid State Physics (ISSP), University of Tokyo, Kashiwa, Chiba 277-8581, Japan}
\author{Hsin Lin}
\affiliation{Centre for Advanced 2D Materials and Graphene Research Centre National University of Singapore, 6 Science Drive 2, Singapore 117546}
\affiliation{Department of Physics, National University of Singapore, 2 Science Drive 3, Singapore 117542}

\author{Shuang Jia}
\affiliation{International Center for Quantum Materials, School of Physics, Peking University, China}\affiliation{Collaborative Innovation Center of Quantum Matter, Beijing,100871, China}

\author{M. Zahid Hasan\footnote{Corresponding author: mzhasan@princeton.edu}}\affiliation {Laboratory for Topological Quantum Matter and Spectroscopy (B7), Department of Physics, Princeton University, Princeton, New Jersey 08544, USA}
\affiliation{Princeton Center for Complex Materials, Princeton Institute for the Science and Technology of Materials, Princeton University, Princeton, New Jersey 08544, USA}

\begin{abstract}
A Weyl semimetal is a new state of matter that hosts Weyl fermions as quasiparticle excitations. The Weyl fermions at zero energy correspond to points of bulk band degeneracy, Weyl nodes, which are separated in momentum space and are connected only through the crystal's boundary by an exotic Fermi arc surface state. We experimentally measure the spin polarization of the Fermi arcs in the first experimentally discovered Weyl semimetal TaAs. Our spin data, for the first time, reveal that the Fermi arcs' spin polarization magnitude is as large as 80\% and lies completely in the plane of the surface. Moreover, we demonstrate that the chirality of the Weyl nodes in TaAs cannot be inferred by the spin texture of the Fermi arcs. The observed non-degenerate property of the Fermi arcs is important for the establishment of its exact topological nature, which reveal that spins on the arc form a novel type of 2D matter. Additionally, the nearly full spin polarization we observed ($\sim80\%$) may be useful in spintronic applications.

\end{abstract}

\date{\today}
\maketitle

The Weyl semimetal can potentially open a new era in condensed matter physics and materials science as it provides the first realization of Weyl fermions, broadens the classification of topological phases beyond insulators, demonstrates exotic quantum anomalies and exhibits novel Fermi arc surface states \cite{Weyl, herring_accidental_1937, abrikosov_properties_1971, nielsen1983adler, Volovik_2003,Murakami2007,Wan2011,Semimetal_Jia,MIT_Weyl,Balents_viewpoint,Burkov2011,Hasan_Na3Bi,Huang2015,Weng2015,Hasan_TaAs,TaAs_Ding,TaAs_Ding_2,NbAs_Hasan,TaP_Hasan,NbP_Hasan}. A Weyl semimetal has a band structure with band crossings between two non-degenerate bands, Weyl nodes, which are each associated with a quantized chiral charge \cite{Wan2011}. It can be understood as a monopole or an anti-monopoles of the Berry curvature in momentum space. Remarkably, the protection of the Weyl fermions does not require any symmetry within band theory. Weyl semimetals may exhibit exceptionally high electron mobilities \cite{Semimetal_Jia,NbP_Transport_1} and may be used to improve electronics by carrying electric currents more efficiently. Moreover, the presence of parallel electrical and magnetic fields can break the apparent conservation of the chiral charge due to the chiral anomaly, making a Weyl metal, unlike ordinary nonmagnetic metals, more conductive with an increasing magnetic field. Furthermore, the Weyl nodes are separated in momentum space and are connected only through the crystal's boundary by a topological surface state, a Fermi arc. Such a band structure would violate quantum mechanics in any purely 2D electron system but it is allowed on the surface of a Weyl semimetal due to the existence of Weyl nodes. These phenomena make new physics accessible and suggest potential applications \cite{Hosur,Excitonic,Axion,Ojanen,CDW,Ashvin_SdH,Weyl-SC-2,Ran_Photon,Weyl_Mott}.

Very recently, the first Weyl semimetal has been experimentally discovered in an inversion breaking, single-crystalline compound TaAs\cite{Hasan_TaAs}. Both the Weyl fermions and the Fermi arcs have been directly observed by photoemission spectroscopy. Here, we study the spin polarization properties of the topological Fermi arc surface states in TaAs \cite{Hasan_TaAs}. Historically, the surface spin texture has played a crucial role for topological insulators because it reveals the $\pi$ Berry phase that demonstrates the topological invariance $\nu_0=1$ \cite{Hasan2010, TI_spin, TI_spin_2}. Hence, it is important to understand the topological meaning of the Fermi arcs' spin texture in a Weyl semimetal. We use our spin data and calculation of the Fermi arcs in TaAs to study this important topic.

Spin-resolved angle-resolved photoemission spectroscopy (SR-ARPES) measurements were at the Institute for Solid State Physics (ISSP) at the University of Tokyo. Photoelectrons were excited by an ultraviolet laser ($h\nu=6.994$ eV). The spin polarization was detected by the very-low-energy electron diffraction (VLEED) spin detectors using pre-oxidized Fe($001$)-p($1\times1$)-O targets. The two spin detectors were placed at an angle of $90^{\circ}$ and were directly attached to a VG-Scienta D80 analyser, enabling simultaneous spin-resolved ARPES measurements for all three spin components as well as high-resolution spin-integrated ARPES experiments. The energy and angle resolutions were set to be better than 20 meV and $0.7^{\circ}$ for the SR-ARPES measurements. Samples were measured at temperature about 20 K and under a vacuum condition better than $1\times10^{-10}$ torr. First-principles calculations were performed by the OPENMX code based on norm-conserving pseudopotentials generated with multi-reference energies and optimized pseudoatomic basis functions within the framework of the generalized gradient approximation (GGA) of density functional theory (DFT) \cite{Perdew}. Spin-orbit coupling was incorporated through $j$-dependent pseudo-potentials. For each Ta atom, three, two, two, and one optimized radial functions were allocated for the $s$, $p$, $d$, and $f$ orbitals ($s3p2d2f1$), respectively, with a cutoff radius of $7$ Bohr. For each As atom, $s3p3d3f2$ was adopted with a cutoff radius of $9$ Bohr. A regular mesh of $1000$ Ry in real space was used for the numerical integrations and for the solution of the Poisson equation. A $k$ point mesh of $17\times17\times5$ for the conventional unit cell was used and experimental lattice parameters \cite{TaAs_Crystal_1} were adopted in the calculations. Symmetry-respecting Wannier functions for the As $p$ and Ta $d$ orbitals were constructed without performing the procedure for maximizing localization and a real-space tight-binding Hamiltonian was obtained \cite{Weng}. This Wannier function based tight-binding model was used to obtain the surface states by constructing a slab with $80$-atomic-layer thickness with Ta on the top and As on the bottom. The surface state band structure was calculated by the surface Green's function technique, which computes the spectral weight near the surface of a semi-infinite system. 

\begin{figure*}
\centering
\includegraphics[width=17cm]{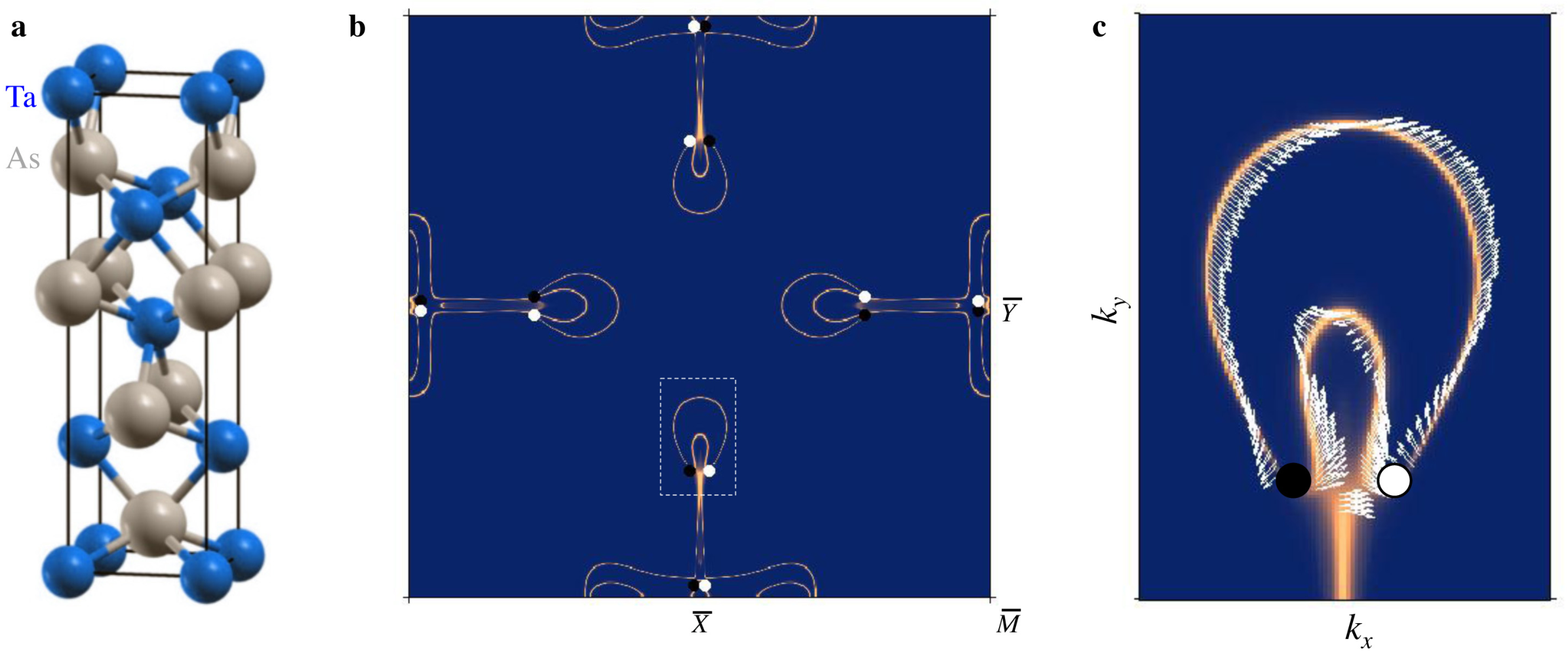}
\caption{\label{Lattice}\textbf{Theoretically calculated surface band structure and spin texture.} (\textbf{a}) Body-centered tetragonal structure of TaAs, shown as stacks of Ta (blue) and As (silver) layers.  The screw-like pattern along the $z$-direction leads to a non-symmporphic C4 rotation symmetry that includes a translation along the $z$-direction by $c/4$. The lattice of TaAs lacks space inversion symmetry. (\textbf{b}) First-principles band structure calculation of the (001) surfaces states of TaAs. The black and white circles indicate the projected Weyl nodes with opposite chirality. (\textbf{c}) Corresponding theoretical spin texture of Fermi arc surface states. The $k$-space range is defined by the white dotted box in (b). }
\end{figure*}

Let us first discuss the essential aspects of the surface state band structure that are important for our investigation of the spin texture. TaAs crystalizes in a body-centered tetragonal lattice system with the space group of $I4_1md$ ($\#109$). Systematic details of the band structures can be found in Ref. \cite{Hasan_TaAs}. Figure~\ref{Lattice}\textbf{a} shows a schematic illustration of TaAs's crystal lattice. It can be seen that the lattice lacks any space-inversion center, which is the key to realizing the Weyl semimetal state \cite{Murakami2007}. First principles calculations showed 24 Weyl nodes in the bulk Brillouin zone (BZ). On the (001) surface of the TaAs, the 24 Weyl nodes project onto 16 points. Eight projected Weyl nodes near the surface BZ boundary ($\bar{X}$ and $\bar{Y}$ points) have a projected chiral charge of $\pm1$. The other eight projected Weyl nodes close to midpoints of the $\bar{\Gamma}-\bar{X}(\bar{Y})$ lines have a projected chiral charge of $\pm2$. They are shown by black and white dots in Fig.~\ref{Lattice}\textbf{b}. The calculated surface state Fermi surface is in excellent agreement with the ARPES data in Ref. \cite{Hasan_TaAs}. Specifically, we identify three main features, namely a bowtie-shaped contour at the $\bar{X}$ point, an elliptical contour at the $\bar{Y}$ point, and a crescent-shaped feature near the midpoint of each $\bar{\Gamma}-\bar{X}(\bar{Y})$ line. Our investigation focuses on  the crescent-shaped feature that consists of two curves (Fermi arcs) that join each other at the two end points, which correspond to projected Weyl nodes with projected chiral charge of $\pm2$. The Fermi arcs near the $\bar{X}$ and $\bar{Y}$ points were not well-resolved in ARPES due to the close proximity of the corresponding Weyl nodes \cite{Hasan_TaAs}. Hence we only focus on the crescent Fermi arcs near the midpoint of each $\bar{\Gamma}-\bar{X}(\bar{Y})$ line. 

\begin{figure*}
\centering
\includegraphics[width=17cm]{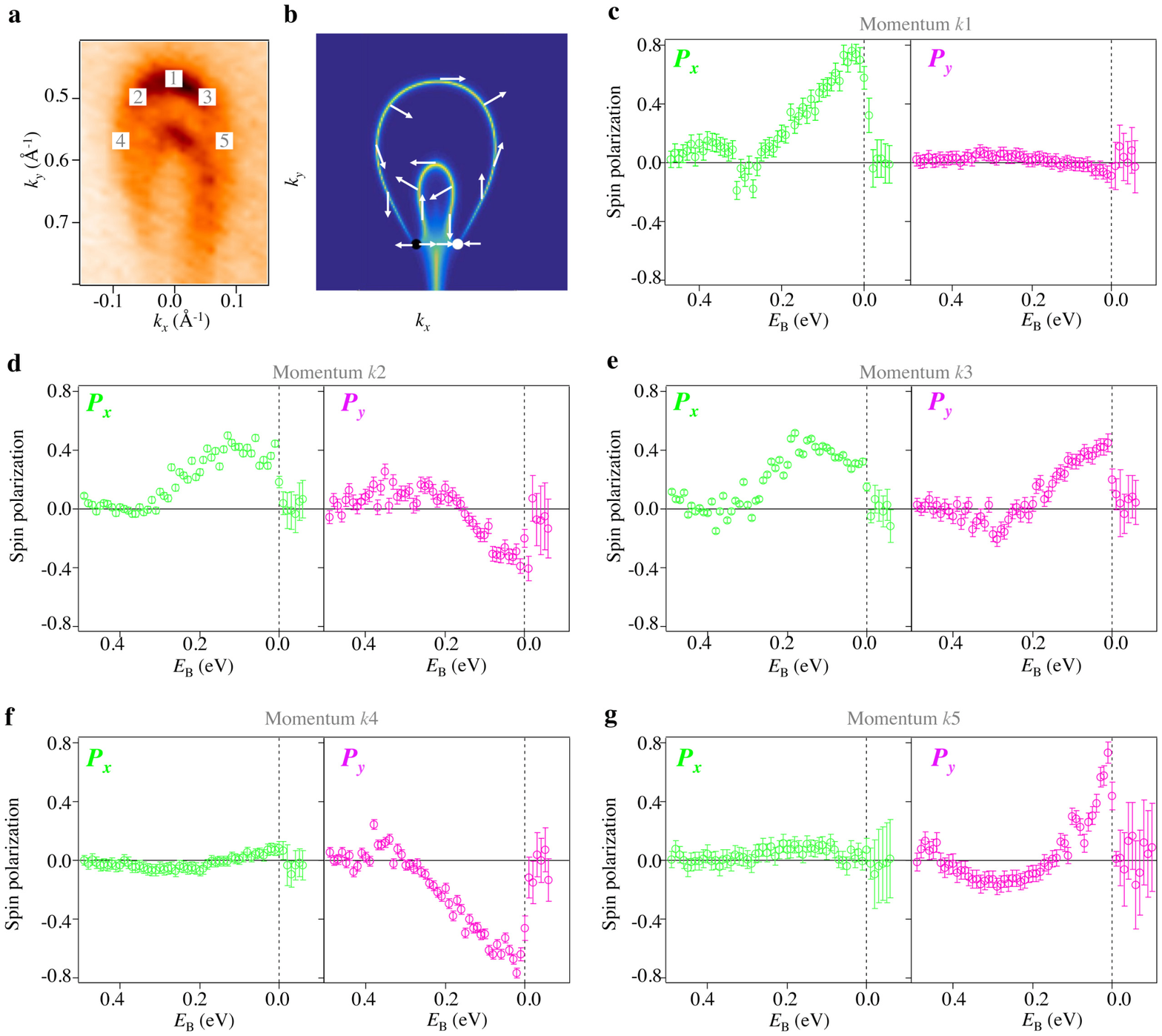}
\caption{\label{Outer}\textbf{Spin texture measurement of outer Fermi arc.} (\textbf{a}) High-resolution ARPES Fermi surface map of crescent Fermi arcs measured with incident photon energy of 90 eV. Numbers 1, 2, 3, 4, and 5 indicate the $k$ locations where spin-resolved ARPES measurements were performed on the outer arc. (\textbf{b}) Schematic illustration of the spin texture. The arrows show the direction of spin polarization on selective $k$ points on the Fermi arcs based on our calculations. (\textbf{c})-(\textbf{g}) Spin-resolved ARPES measurements at 1, 2, 3, 4 and 5 correspond to panels denoted by $k_1$, $k_2$,  $k_3$, $k_4$, and $k_5$, respectively. Shown in panels (\textbf{c}-\textbf{g}) is the measured in-plane spin polarization for points 1, 2, 3, 4, and 5. The spin-polarization along the $\hat{x}$ and  $\hat{y}$ are colored green and pink, respectively. The measured in-plane spin texture is consistent with the theoretically calculated spin-texture as shown in (b).}
\end{figure*}

Figure~\ref{Lattice}\textbf{c} shows the calculated spin texture of the crescent Fermi arcs. Denoted by white arrows is the direction of spin polarization along the Fermi arcs. The direction of spin polarization rotates clockwise as one travels around the outer Fermi arc in a counter-clockwise fashion. If one continues along this trajectory, but now starting from the previous end-point and travel along the inner arc in an anti-clockwise fashion, we observe that the spin direction also rotates clockwise. Therefore, the spin polarization direction for the extrema of the outer and inner Fermi arcs are opposite to each other. An observation to make here is the constraint mirror symmetry enforces on the allowed spin polarization direction. As illustrated in Fig 1c, the crescent Fermi arcs intersect the high-symmetry line ($\bar{\Gamma}-X$), which is invariant under the reflection $\mathcal{M}_y: y ->  -y$. This crescent Fermi arc lies on a high symmetry line that cuts along the middle of both arcs. This mirror symmetry should therefore only allow, for the case of Fig.~\ref{Lattice}\textbf{c}, spin polarization direction along $\pm k_x$, which is consistent with our calculations for the spin texture at extrema of both inner and outer arcs.  In Fig.~\ref{Outer}\textbf{a} we show a high-resolution ARPES Fermi surface map of crescent Fermi arcs that was measured with an incident photon energy of 90 eV. Labeled on Fig.~\ref{Outer}\textbf{a} are the $k_i$ points ($i=1, 2, 3, 4, 5$) on the outer Fermi arc that corresponds to the spin polarization direction calculated,shown in Fig.~\ref{Outer}\textbf{b}, and experimentally measured, shown in Figs.~\ref{Outer}\textbf{c-g}. The spin polarization direction at $k_1$, the extremum of the outer arc, was measured to be nearly $80\%$ along the $+k_x$ direction and approximately $0\%$ spin polarized along the  $\pm k_y$ direction, which is consistent with our calculations in Fig.~\ref{Outer}\textbf{b}. In addition, the measured spin polarization direction at $k_2$ and $k_3$ shows $~40\%$ spin polarization along $+k_x$ and  $-k_y$ direction and $~40\%$ spin polarization along $+k_x$ and  $+k_y$ direction, respectively. Finally, the measured spin polarization direction for the outer Fermi arc at $k_4$ and $k_5$, shows $~0\%$ spin polarization along $\pm k_x$ and $~80\%$ in the $-k_y$ direction and  $+k_y$ direction, respectively. By comparing the in-plane spin polarization direction defined in Figs.~\ref{Outer}\textbf{c-g} with spin texture calculated and displayed in Fig.~\ref{Outer}\textbf{b}, it becomes apparent that our experimental results are consistent with our theoretical predictions.  

\begin{figure*}
\centering
\includegraphics[width=17cm]{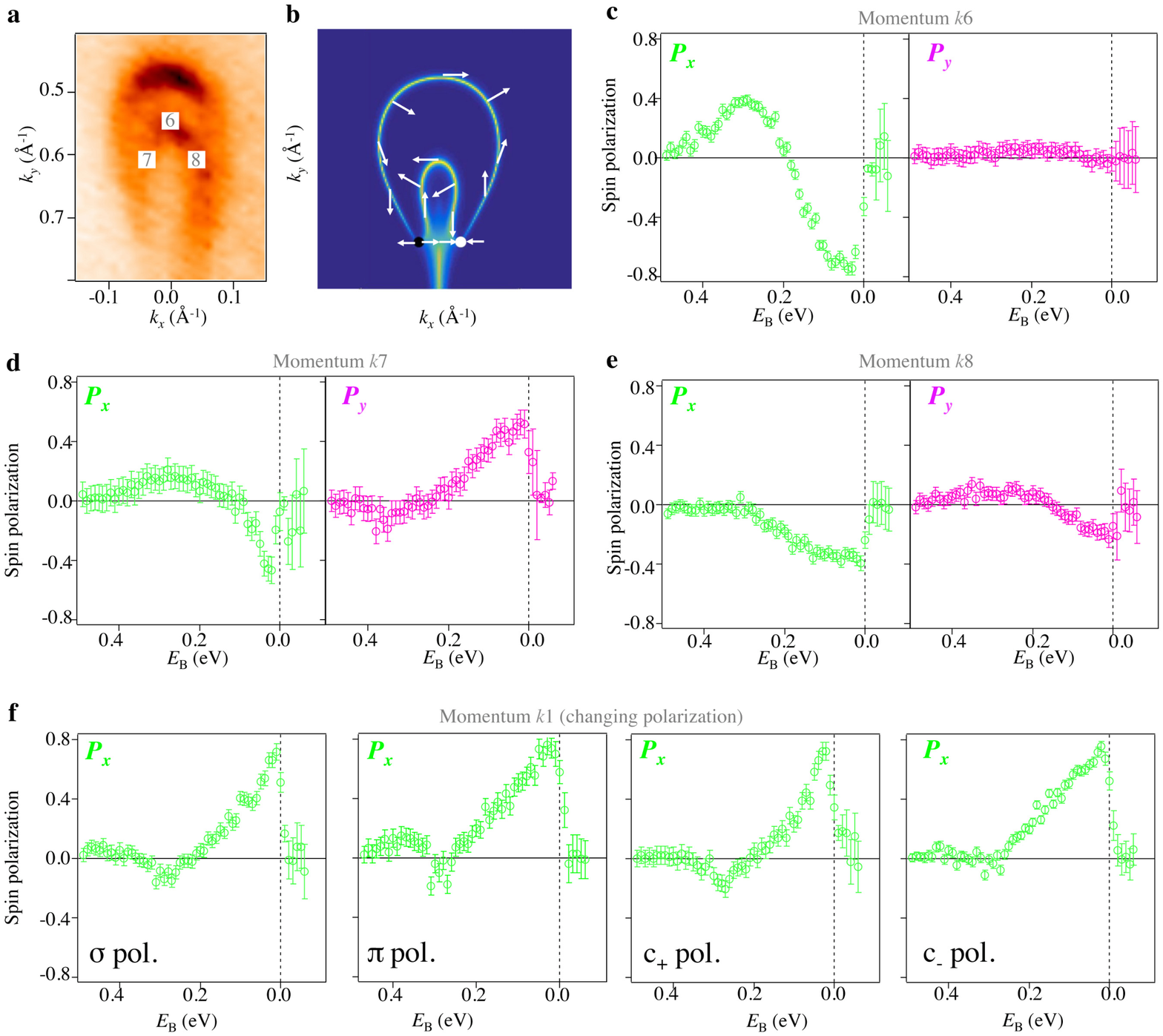}
\caption{\label{Inner}\textbf{Spin texture measurement of inner Fermi arc.} (\textbf{a}) High-resolution ARPES Fermi surface map of crescent Fermi arcs measured with incident photon energy of 90 eV. Numbers 6, 7, and 8 indicate the $k$ locations where spin-resolved ARPES measurements were performed on the inner arc. (\textbf{b}) Schematic illustration of the spin texture. The arrows show the direction of spin polarization on selective $k$ points on the Fermi arcs based on our calculations. (\textbf{c})-(\textbf{g}) Spin-resolved ARPES measurements at 6-8 correspond to panels denoted by $k_6$, $k_7$, and $k_8$, respectively. Shown in panels (\textbf{c}-\textbf{g}) is the measured in-plane spin polarization for points 6-8. The spin-polarization along the $\hat{x}$ and  $\hat{y}$ are colored green and pink, respectively. The measured in-plane spin texture is consistent with the theoretically calculated spin-texture as shown in (b). (\textbf{f}) Spin polarization measurements at $k_1$ (defined in Fig. 2\textbf{a}). The four datasets shown in this panel were performed under identical conditions except that the incident light polarization was varied. We observe that the spin polarization does not depend on the photon polarization, which excludes the possibility that our data is due to final state effects \cite{Rader}.}
\end{figure*}

Lets now perform a similar spin texture analysis on the inner crescent Fermi arc. In Fig.~\ref{Inner}\textbf{a} we present the same Fermi surface map of the Fermi arcs presented in Fig.~\ref{Outer}\textbf{a}, but now the inner Fermi arc is labeled with $k_i$ points ($i=6, 7, 8$). A schematic illustration of the spin texture predicted by calculation is shown in Fig.~\ref{Inner}\textbf{b}, with the spin polarization direction denoted by white arrows. The spin polarization direction at $k_6$, the extremum of the inner arc, was measured to be nearly $80\%$ along the $-k_x$ direction and approximately $0\%$ spin polarized along the  $\pm k_y$ direction, which is consistent with our calculations in Fig.~\ref{Inner}\textbf{b}. Now, the measured spin polarization direction at $k_7$ and $k_8$ shows $~45\%$ spin polarization along $-k_x$ and  $+k_y$ direction and $~30\%-40\%$ spin polarization along $-k_x$ and  $-k_y$ direction, respectively. Similar to the outer Fermi arc presented in Fig.~\ref{Outer}, the spin texture calculated and exhibited in Fig.~\ref{Outer}\textbf{b} is consistent with our experimental results shown in Figs.~\ref{Inner}\textbf{c-e}. We have also carefully checked the possibility of final state effects to our spin measurements. Specifically, we repeat the same spin polarization under different polarizations of the incident light. Our data (Fig.~\ref{Inner}\textbf{f}) show that the spin polarization does not change as a function of polarization, which excludes the final state effects \cite{Rader}.

\begin{figure*}
\centering
\includegraphics[width=15cm]{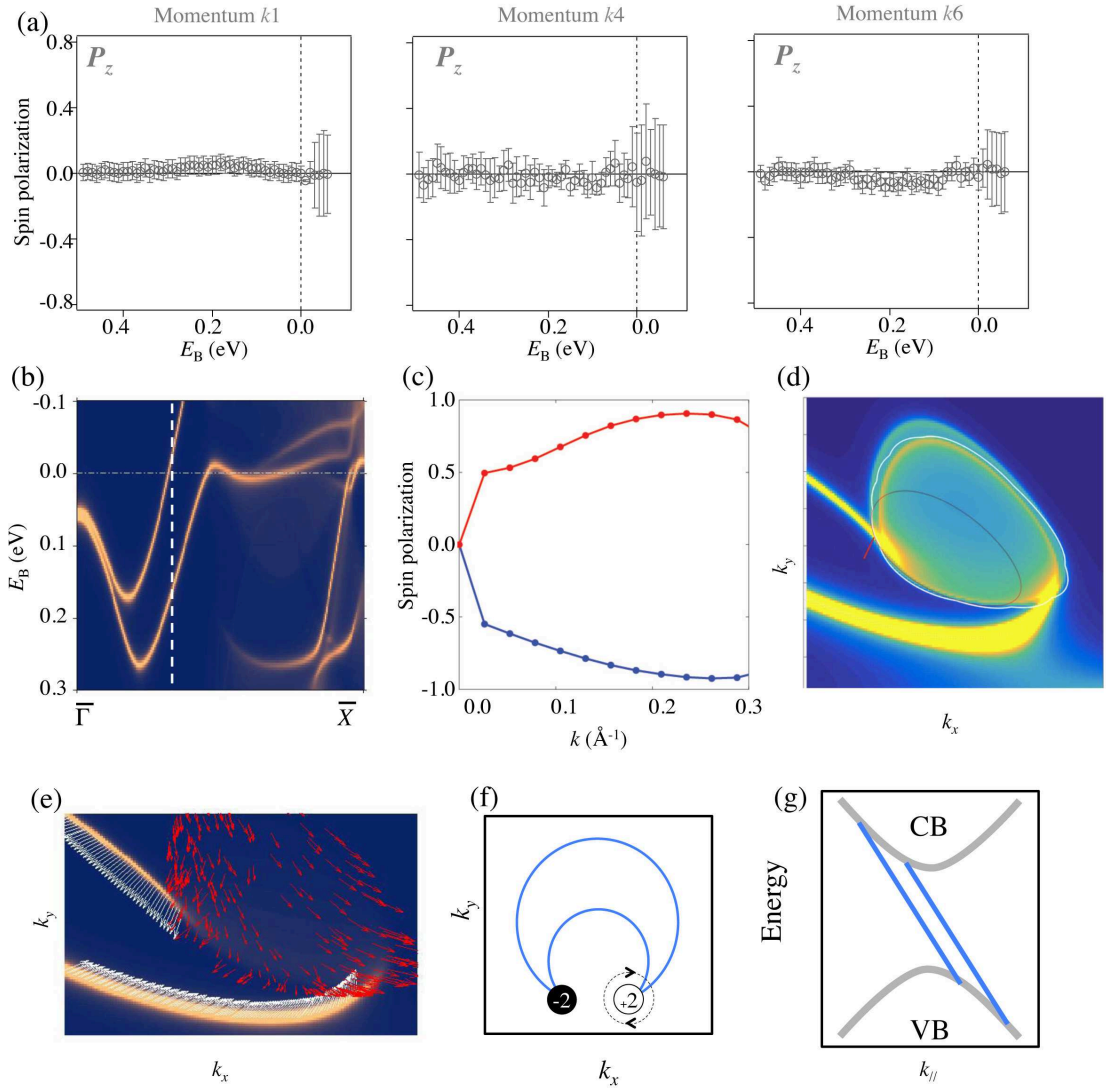}
\caption{\label{THY}\textbf{Lack of out-of-plane spin polarization, large spin polarization magnitude, and deviation between spin textures of the Fermi arc and Weyl cones.} (\textbf{a}) Spin polarization measurements along the out-of-plane direction for $k_1$, $k_4$, and $k_6$. No out-of plane spin polarization is measured within in our experimental resolution. (\textbf{b}) Theoretically calculated surface energy dispersion along $\bar{\Gamma}-\bar{X}$. (\textbf{c}) The magnitude of spin polarization for the two bands that form a Kramers' pair near the Fermi level at the $\bar{\Gamma}$ point in panel (b). These two bands arise from the two Fermi arcs of one crescent-shaped feature. For each band, we show the magnitude of spin polarization near the $\bar{\Gamma}$ along the $\bar{\Gamma}$ to $\bar{X}$. In other words, the range of the $x$ axis is from the $\bar\Gamma$ point to the white dotted lines in panel (b).}
\end{figure*}

The out-of-plane spin polarization data are shown in Fig.~\ref{THY}\textbf{a}. From the data, it is clear that no out-of plane spin ($P_z$) polarization is measured within in our experimental resolution. This is consistent with our calculations, where also zero out-of plane spin polarization is found everywhere in the surface BZ. The lack of out-of plane spin polarization is in fact guaranteed by symmetries, namely, time reversal symmetry $\mathcal{T}$ and two-fold rotational symmetry along the $\hat{z}$ axis $C_{2z}$. Specifically, on the surface, $\mathcal{T}$ and $C_{2z}$ have the same effect, bringing $(k_x,k_y)$ to $(-k_x,-k_y)$. Start with the spin at $(k_x,k_y)$ as $(S_x,S_y,S_z)$, we could get the surface spin at $(-k_x,-k_y)$ by $\mathcal{T}(S_x,S_y,S_z)=(-S_x,-S_y,-S_z)$ or by  $C_{2z}(S_x,S_y,S_z)=(-S_x,-S_y,S_z)$. However, we know that these two symmetries should lead to the same spin at $(-k_x,-k_y)$. This is because one has $|\psi (k_x,k_y) \rangle  =  \mathcal{T}C_{2z} |\psi (k_x,k_y) \rangle$ up to some phase factor, which is a consequence of $(\mathcal{T}C_{2z})^2 = (\mathcal{T}^2) (C_{2z}^2) = (-1)(-1) = +1$. Therefore, we have $S_z=-S_z=0$.

We also study the origin of the observed large spin polarization. We check the magnitude of the spin polarization in theoretical calculations. As shown in Fig.~\ref{THY}\textbf{b}, the two bands that forms a Kramers' pair near the Fermi level at the $\bar{\Gamma}$ point are the two crescent Fermi arcs. The results in Fig.~\ref{THY}\textbf{b} show that the two Fermi arcs have opposite spin polarization direction and a polarization magnitude of about $85\%$. These results are consistent with the experimental data (Figs.~\ref{Outer} and~\ref{Inner}). To understand the large polarization, we decompose the total spin texture into contributions from different orbitals. In a spin-orbit coupled system, the physical spin is not the eigenstate of the system and each orbital has its own spin texture. The total spin texture is the sum of all spin textures that arise from different orbitals. In the case of TaAs, our calculation shows that the Fermi arcs arise the As $p_x$ orbital and the Ta $d_{xz}$ orbital. The spin textures of these two orbitals sum up in a constructive way. Therefore, the total spin polarization is close to unit 1. On the other hand, it is well known that the spin polarization of Bi$_2$Se$_3$ surface states \cite{TI_spin_2} is only about $40\%$. This is because the spin textures of $p_x$, $p_y$, and $p_z$ orbitals add up destructively. 

Finally, we investigate the topological ``information'' of the Fermi arc. We first briefly review the role of the non-spin part of the band structure. As elaborated in Refs. \cite{Hasan_TaAs, NbP_Hasan, TaP_Hasan}, one can prove for the Fermi arcs by resolving the sign of the Fermi velocity of the surface bands. Specifically, if we choose the dotted circle in Fig.~\ref{THY}\textbf{f} that encloses a $+2$ projected Weyl node. The dotted circle can be viewed as the projection of a circular pipe that goes through the bulk BZ along the $k_z$ direction. Such a pipe is a closed manifold. We know that the bulk band structure has to be fully gapped on this pipe as the bulk band gap only vanishes at the Weyl nodes. A Chern number is well-defined for a fully gapped band structure and the Chern number of the pipe must be $+2$ as the pipe encloses a net chiral charge of $+2$. Therefore, the boundary of the pipe, i.e., the dotted circle on the surface, must have two chiral modes, as shown in Fig.~\ref{THY}\textbf{g}. Hence one can prove that the two curves in Fig.~\ref{THY}\textbf{f} are two Fermi arcs by showing that they have the same sign for their Fermi velocities. This tells us that one can already prove the existence of Fermi arcs with the non-spin part of the band structure. As for the spin texture, the non-degenerate (spin polarized) property is important. If the surface bands were doubly degenerate, the chiral modes in Fig.~\ref{THY}\textbf{g} would give rise to a Chern number of $+4$, which is inconsistent with the chiral charge of $+2$ in the bulk. However,  the non-degenerate property represents only a small fraction of the information from the Fermi arc spin texture. An obvious question to ask is that whether the texture, i.e., the $k$ space configuration of the spin polarization, also carries any topological meaning. One proposal is that one can infer the chirality of bulk Weyl cone by studying the spin texture of the Fermi arcs at the $k$ points where they merge with each other. However, we note that this proposal is not feasible in TaAs for the following reasons. First, a significant part of the bulk band projection arises from other irrelevant (non-Weyl) bulk Fermi surfaces (Fig.~\ref{THY}\textbf{d}). Second, the spin polarization of the Fermi arc does not match with that of the projected bulk bands at $k$ points where they merge (Fig.~\ref{THY}\textbf{e}). This is the $k$ points where they merge We expect surface resonance states at $k$ points where surface states enter the bulk. Hence the surface spin should slowly evolve and only become fully consistent with the bulk when it goes deeply into the bulk projection as the wavefunction become completely delocalized from the surface. Third, as can be seen from the red arrows in Fig.~\ref{THY}\textbf{e}, the bulk spin texture does not follow the monopole/antimonopole configuration. In fact, the Weyl nodes are monopoles of Berry curvature, and the Berry curvature is in general different from the physical spin in a spin-orbit coupled system. Hence even if one were able to infer the bulk spin texture from the surface spin texture, it cannot prove the monopole property in the case of TaAs. Nevertheless, our observation provides a new type of spin texture that arises from a new type of 2D electron gas, the Fermi arc surface states. The spin texture is completely in-plane. The nearly full spin polarization is favorable for spin applications based on the Fermi arcs of TaAs.


\clearpage

\clearpage
\clearpage

\clearpage
\addtocounter{figure}{-1}
\begin{figure*}[t!]
\caption{(\textbf{d}) A zoomed-in calculation near a projected Weyl node slightly above the Fermi level. The two Fermi arcs are seen to merge into the bulk band projection, which is noted by the white contour. The black contour shows the projection of the Weyl cones only. We see that the total projection (white contour) is larger than the Weyl cone projection (black contour). This demonstrates that there are additional irrelevant (non-Weyl) bulk Fermi surface contributing to the bulk projection. (\textbf{e}) The spin-texture calculation for (d), which shows the spin-polarization of the Fermi arc surface states with white arrows, and red arrows for the bulk pocket around the Weyl node.  (\textbf{f}) A schematic showing crescent shaped Fermi arcs (blue curve) connecting to projected Weyl nodes. (\textbf{g}) An illustration of the band structure along the black dotted circle.}
\end{figure*}

\end{document}